\documentclass[twocolumn,showpacs,superscriptaddress]{revtex4-1}
\usepackage{graphicx}
\usepackage{units}
\usepackage{multirow}
\usepackage{bigstrut}
\usepackage{overpic}
\usepackage{array}
\usepackage{color}
\usepackage{amsmath}
\RequirePackage{lineno}

\usepackage[perpage,symbol]{footmisc}

\newcommand{\gevc}{\,\unit{GeV}/c}

\begin{document}

\modulolinenumbers[2]

\setlength{\oddsidemargin}{-0.5cm} \addtolength{\topmargin}{15mm}

\title{\boldmath \large Observation of $\Lambda^+_{c}\to nK^0_S\pi^+$ }

\author{
  \small
      M.~Ablikim$^{1}$, M.~N.~Achasov$^{9,e}$, S.~Ahmed$^{14}$,
      X.~C.~Ai$^{1}$, O.~Albayrak$^{5}$, M.~Albrecht$^{4}$,
      D.~J.~Ambrose$^{44}$, A.~Amoroso$^{49A,49C}$, F.~F.~An$^{1}$,
      Q.~An$^{46,a}$
% ...
      \\
      \vspace{0.2cm}
      (BESIII Collaboration)\\
      \vspace{0.2cm} {\it
        $^{1}$ Institute of High Energy Physics, Beijing 100049, People's Republic of China\\
% ...
$^{4}$ Bochum Ruhr-University, D-44780 Bochum, Germany\\
        $^{5}$ Carnegie Mellon University, Pittsburgh, Pennsylvania 15213, USA\\
% ...
$^{9}$ G.I. Budker Institute of Nuclear Physics SB RAS (BINP), Novosibirsk 630090, Russia\\
% ...
$^{14}$ Helmholtz Institute Mainz, Johann-Joachim-Becher-Weg 45, D-55099 Mainz, Germany\\
% ...
$^{44}$ University of Rochester, Rochester, New York 14627, USA\\
% ...
$^{46}$ University of Science and Technology of China, Hefei 230026, People's Republic of China\\
% ...
$^{49}$ (A)University of Turin, I-10125, Turin, Italy; (B)University of Eastern Piedmont, I-15121, Alessandria, Italy; (C)INFN, I-10125, Turin, Italy\\
% ...
$^{a}$ Also at State Key Laboratory of Particle Detection and Electronics, Beijing 100049, Hefei 230026, People's Republic of China\\
% ...
$^{e}$ Also at the Novosibirsk State University, Novosibirsk, 630090, Russia\\
% ...
    \vspace{0.4cm}
}
}

\begin{abstract}
We report the first direct measurement of decays of the $\Lambda^+_c$
baryon involving the neutron. The analysis is performed using 567
pb$^{-1}$ of $e^+e^-$ collision data collected at $\sqrt{s}=4.599$ GeV with
the BESIII detector at the BEPCII collider. We observe the decay $\Lambda^+_{c}\rightarrow n K^0_S\pi^+$ and measure
the absolute branching fraction to be
$\mathcal{B}({\Lambda^+_{c}\rightarrow n K^0_S\pi^+})=(1.82\pm0.23({\rm stat})\pm0.11({\rm syst}))\%$.
A comparison to $\mathcal{B}({\Lambda^+_{c}\rightarrow p(\bar{K}\pi)^0})$
provides an important test of isospin symmetry and final state interactions.
\end{abstract}


\maketitle


The ground-state charmed baryon $\Lambda^+_c$ decays eventually into a proton or a neutron, each taking about half of the total branching fraction (BF)~\cite{pdg2014}.
However, to date no direct measurement of the decay modes involving a neutron has been performed.
It has been argued that isospin symmetry works well in the charmed baryon sector~\cite{1601.04241}.
Comparing BFs of the final states with a neutron to the final states with a proton provides an important observable in testing isospin symmetry in $\Lambda^+_c$ three-body decays~\cite{1601.04241}.
The decay $\Lambda_c^+\rightarrow n\bar{K}^0\pi^+$ is the most favored decay of the $\Lambda_c$ involving a neutron.
Under the isospin symmetry, its amplitude is related to those of the most favored proton modes $\Lambda_c^+\rightarrow p K^-\pi^+$ and  $\Lambda_c^+\rightarrow p \bar{K}^0\pi^0$ as
$\mathcal{A}(n\bar{K}^0\pi^+) + \mathcal{A}(p K^-\pi^+) + \sqrt{2}\mathcal{A}(p\bar{K}^0\pi^0) =0$.
Hence, precise measurement of the BF for $\Lambda_c^+\rightarrow n\bar{K}^0\pi^+$ provide stringent test on the isospin symmetry in the charmed baryon decays by examining this triangle relation.

Furthermore, study of $\Lambda_c^+\rightarrow n\bar{K}^0\pi^+$ is important to explore the decay mechanism of the $\Lambda^+_c$, especially the factorization scheme and the involved final state interaction~\cite{1601.04241, Cheng}. In the three-body $\Lambda^+_c$ decay to $N\bar{K}\pi$, the total decay amplitudes can be decomposed into two isospin amplitudes of the $N\bar{K}$ system as isosinglet ($I^{(0)}$) and isospin-one ($I^{(1)}$). In the factorization limit, the color-allowed tree diagram, in which the $\pi^+$ is emitted and the $N\bar{K}$ is an isosinglet, dominates $I^{(0)}$, and $I^{(1)}$ is expected to be small compared to $I^{(0)}$ as it can only proceed through the color-suppressed tree diagrams. Though the factorization scheme is spoiled  in charmed meson decays, whether this scheme is valid in the charmed baryon $\Lambda_c^+$ decays  is of great interest to both theorists and experimentalists and  strongly deserves the experimental investigation. The measurement of BF for $\Lambda_c^+\rightarrow n\bar{K}^0\pi^+$ can validate or falsify this scheme. Together with the $\Lambda_c^+\rightarrow p(\bar{K}\pi)^0$, the $\Lambda_c^+\rightarrow n\bar{K}^0\pi^+$ can be used to determine  the magnitudes of the two isospin amplitudes and their phase difference, which provides crucial information on the final state interaction. In addition, hight statistics data will facilitate to understand the resonant structures~\cite{Miyahara:2015cja, Xie:2016evi} in the three-body $\Lambda_c$ decays and test the SU(3) flavor symmetry~\cite{1601.04241}.
Throughout the paper, charge conjugate modes are always implied.

This Letter reports on the observation of the final states with a neutron $\Lambda_c^+\rightarrow nK^0_S\pi^+$.
The data analyzed correspond to $566.93\pm0.11$ pb$^{-1}$~\cite{lum} of $e^+e^-$ annihilations accumulated with the BESIII experiment at
$\sqrt{s}=4.599$~GeV~\cite{Ablikim:2015zaa}. This energy is slightly above the mass threshold of a
$\Lambda_c^{+}\bar{\Lambda}_c^{-}$ pair, at which $\Lambda_c^{+}\bar{\Lambda}_c^{-}$ are produced in pairs and no additional hadron is kinematically allowed.
The analysis technique in this work, which was first applied in the
Mark~III experiment~\cite{prl62_1821}, is specific for charm hadron pairs produced near threshold.
First, we select a data sample of $\bar{\Lambda}^-_c$ baryons by reconstructing exclusive
hadronic decays, called the single tag (ST) sample. Then, we
search for $\Lambda_c^+\rightarrow nK^0_S\pi^+$ in the system
recoiling against the ST $\bar{\Lambda}^-_c$ baryons, called the double tag (DT) sample.
In the final state $nK^0_S\pi^+$, the neutron is not detected, and
its kinematics is deduced by four-momenta conservation.
The absolute BF of $\Lambda_c^+\rightarrow nK^0_S\pi^+$ is then determined from the
probability of detecting the process $\Lambda_c^+\rightarrow nK^0_S\pi^+$ in the ST sample.
This method provides a clean and straightforward BF
measurement independent of the total number of
$\Lambda^+_c\bar{\Lambda}^-_c$ events produced.

The BESIII detector is a cylindrical detector with
a solid-angle coverage of 93\% of $4\pi$ that operates at the
BEPCII collider. It consists of a Helium-gas based main drift chamber (MDC), a plastic
scintillator time-of-flight (TOF) system, a CsI~(Tl) electromagnetic
calorimeter (EMC), a superconducting solenoid providing a 1.0\,T
magnetic field and a muon counter. The charged particle momentum
resolution is 0.5\% at a transverse momentum of 1\,$\gevc$. The
photon energy resolution in EMC is 2.5\% in the barrel and 5.0\% in the end-caps at energies of 1\,GeV.
More details about the design and performance of the detector are given in
Ref.~\cite{Ablikim:2009aa}.

A GEANT4-based~\cite{geant4} Monte Carlo (MC) simulation package,
which includes a description of the detector geometry and the
detector response, is used to determine the detection efficiency and
to estimate potential backgrounds. Signal MC samples of a
$\Lambda_c^+$ baryon decaying only to $nK^0_S\pi^+$ together with a
$\bar{\Lambda}_c^-$ decaying only to the studied tag modes are
generated by the MC event generator KKMC~\cite{kkmc} using
EVTGEN~\cite{nima462_152}, including the effects of initial-state radiation (ISR)~\cite{SJNP41_466}. Final-state radiation (FSR)
off the charged tracks is simulated with the PHOTOS package~\cite{plb303_163}.
The $\Lambda_c^+\rightarrow nK^0_S\pi^+$ decay is simulated using a phase space model since the two-body invariant mass spectra found in data for
$M_{n\pi^+}$, $M_{nK^0_S}$ and $M_{K^0_S\pi^+}$ show no obvious structure.
To study backgrounds, inclusive MC samples consisting of generic $\Lambda_c^+\bar{\Lambda}_c^-$
events, $D_{(s)}^{*}\bar{D}_{(s)}^{(*)}+X$ production, ISR return to the charmonium(-like)
$\psi$ states at lower masses, and QED processes are generated.
All decay modes of the $\Lambda_c$, $\psi$ and $D_{(s)}$ as
specified in the Particle Data Group (PDG)~\cite{pdg2014} are
simulated by the EVTGEN MC generator, while the unknown decays of the $\psi$
states are generated with LUNDCHARM~\cite{lundcharm}.


The ST $\bar{\Lambda}^-_c$ baryons are reconstructed using eleven hadronic
decay modes as listed in the first column of Table~\ref{tab:deltaE_1}, where the
intermediate particles $K^0_S$, $\bar{\Lambda}$, $\bar{\Sigma}^0$,
$\bar{\Sigma}^-$ and $\pi^0$ are reconstructed through their decays of
$K^0_S\rightarrow \pi^+\pi^-$, $\bar{\Lambda}\rightarrow
\bar{p}\pi^+$, $\bar{\Sigma}^0\rightarrow \gamma\bar{\Lambda}$ with
$\bar{\Lambda}\rightarrow \bar{p}\pi^+$, $\bar{\Sigma}^-\rightarrow
\bar{p}\pi^0$ and $\pi^0\rightarrow \gamma\gamma$, respectively.

Charged tracks are required to have polar angles within
$|\cos\theta|<0.93$, where $\theta$ is the polar angle of the
charged track with respect to the beam direction. Their distances of
closest approach to the interaction point (IP) are required to be
less than 10\,cm along the beam direction and less than 1\,cm in the
perpendicular plane. Tracks originating from $K^0_S$ and $\Lambda$
decays are not subjected to these distance requirements. To
discriminate pions from kaons, the specific ionization energy loss ($dE/dx$) in the MDC and TOF information are
used to obtain particle identification (PID) probabilities for the pion ($\mathcal{L}_{\pi}$) and
kaon ($\mathcal{L}_K$) hypotheses. Pion and kaon candidates are
selected using $\mathcal{L}_{\pi} > \mathcal{L}_{K}$ and
$\mathcal{L}_{K} > \mathcal{L}_{\pi}$, respectively. For proton
identification, information from $dE/dx$, TOF, and EMC are combined
to calculate the PID probability $\mathcal{L'}$, and a charged track
satisfying $\mathcal{L'}_p> \mathcal{L'}_{\pi}$ and
$\mathcal{L'}_{p} > \mathcal{L'}_{K}$ is identified as a proton
candidate.

Photon candidates are reconstructed from isolated clusters in the
EMC in the regions $|\cos\theta| \le 0.80$ (barrel) and $0.86 \le
|\cos\theta|
  \le 0.92$ (end cap). The deposited energy of a neutral cluster is required to be larger than 25 (50) MeV in
barrel(end cap) region, and the angle between the photon candidate
and the nearest charged track must be larger than 10$^\circ$. To
suppress electronic noise and energy deposits unrelated to the
events, the difference between the EMC time and the event start time
is required to be within (0, 700)~ns. To reconstruct $\pi^0$
candidates, the invariant mass of the accepted photon pair is
required to be within $(0.110,~0.155)$~GeV$/c^2$. A kinematic fit is
performed to constrain the $\gamma\gamma$ invariant mass to the
nominal $\pi^0$ mass~\cite{pdg2014}, and the $\chi^2$ of the
kinematic fit is required to be less than 20. The fitted momenta of
the $\pi^0$ are used in the further analysis.

To reconstruct $K^0_S$ and $\bar{\Lambda}$ candidates, a vertex-constrained fit
is applied to $\pi^+\pi^-$ and $\bar{p}\pi^+$ combinations, and the fitted track parameters are used in the further analysis.
The signed decay length $L$ of the secondary vertex to the IP is also required to be larger than zero.
The same PID requirements as mentioned before are applied to the proton candidate, but not to the $\pi$ candidate.
The invariant masses $M_{\pi^+\pi^-}$, $M_{\bar{p}\pi^+}$,
$M_{\gamma\bar{\Lambda}}$ and $M_{\bar{p}\pi^0}$ are required to be
within $(0.485,~0.510)$~GeV/$c^2$, $(1.110,~1.121)$~GeV/$c^2$,
$(1.179,~1.205)$~GeV/$c^2$ and $(1.173,~1.200)$~GeV/$c^2$ to select
candidates for $K^0_S$, $\bar{\Lambda}$, $\bar{\Sigma}^0$ and
$\bar{\Sigma}^-$ candidates, respectively.

For the ST mode $\bar{p}K^0_S\pi^0$, the backgrounds involving $\bar{\Lambda}$ and
$\bar{\Sigma}^-$ are rejected by rejecting any event
with $M_{\bar{p}\pi^+}\in (1.105,1.125)$~GeV/$c^2$ and
$M_{\bar{p}\pi^0}\in (1.173, 1.200)$~GeV/$c^2$.
For the ST modes of $\bar{\Lambda}\pi^+\pi^-\pi^-$ and
$\bar{\Sigma}^-\pi^+\pi^-$, the backgrounds involving $K^0_S$ and $\Lambda$ as intermediate states are suppressed by requiring $M_{\pi^+\pi^-}\notin (0.480, 0.520)$~GeV/$c^2$
and $M_{\bar{p}\pi^+}\notin (1.105, 1.125)$~GeV/$c^2$.

\begin{table}[tp!]
\caption{ ST modes, $\Delta E$ requirements and  ST yields
$N_{\bar{\Lambda}_c^-}$ in data. The errors are statistical only.}
\begin{center}
\begin{tabular}
{llc} \hline \hline Mode~&~~~$\Delta E$~(GeV)~~~&$N_{\bar{\Lambda}_c^-}$ \\
\hline
 $\bar{p} K^0_S$                & [$-$0.025, 0.028] &   $1066\pm33$  \\
 $\bar{p} K^+\pi^-$             & [$-$0.019, 0.023] &   $5692\pm88$  \\
 $\bar{p}K^0_S\pi^0$            & [$-$0.035, 0.049] &  ~~$593\pm41$  \\
 $\bar{p} K^+\pi^-\pi^0$        & [$-$0.044, 0.052] &   $1547\pm61$  \\
 $\bar{p} K^0_S\pi^+\pi^-$      & [$-$0.029, 0.032] &  ~~$516\pm34$  \\
 $\bar{\Lambda}\pi^-$           & [$-$0.033, 0.035] &  ~~$593\pm25$  \\
 $\bar{\Lambda}\pi^-\pi^0$      & [$-$0.037, 0.052] &   $1864\pm56$  \\
 $\bar{\Lambda}\pi^-\pi^+\pi^-$ & [$-$0.028, 0.030] &  ~~$674\pm36$  \\
 $\bar{\Sigma}^0\pi^-$          & [$-$0.029, 0.032] &  ~~$532\pm30$  \\
 $\bar{\Sigma}^-\pi^0$          & [$-$0.038, 0.062] &  ~~$329\pm28$  \\
 $\bar{\Sigma}^-\pi^+\pi^-$     & [$-$0.049, 0.054] &   $1009\pm57$  \\
 \hline
 All tags &   & $14415\pm159$ \\
\hline \hline
\end{tabular}
\label{tab:deltaE_1}
\end{center}
\end{table}

The ST $\bar{\Lambda}^-_c$ signal candidates are identified using the variable of beam
constrained mass, $M_{\rm BC}\cdot c^2 \equiv \sqrt{E^2_{\rm
beam}-|\overrightarrow{p}_{\bar{\Lambda}^-_c}\cdot c|^2}$, where $E_{\rm
beam}$ is the beam energy and
$\overrightarrow{p}_{\bar{\Lambda}^-_c}$ is the momentum of the
$\bar{\Lambda}^-_c$ candidate. To improve the signal purity, the
energy difference $\Delta E \equiv E_{\rm beam}-E_{\bar{\Lambda}^-_c}$ for
each candidate is required to be within approximately
$\pm3\sigma_{\Delta E}$ around the $\Delta E$ peak, where
$\sigma_{\Delta E}$ is the $\Delta E$ resolution and
$E_{\bar{\Lambda}^-_c}$ is the reconstructed $\bar{\Lambda}^-_c$
energy. The explicit $\Delta E$ requirements for the different modes are
listed in Table~\ref{tab:deltaE_1}. The yield of each tag mode is
obtained from fits to the $M_{\rm BC}$ distributions in the signal
region $(2.280, 2.296)$~GeV/$c^2$, which is the same as in Ref.~\cite{bes3lamev}. The yields of
reconstructed singly tagged $\bar{\Lambda}_c^-$ baryons are listed
in Table~\ref{tab:deltaE_1}. Finally, we obtain the total ST yield
summed over all 11 modes to be $N^{\rm
tot}_{\bar{\Lambda}^-_c}=14415\pm159$, where the error is statistical only.

Candidates for the decay $\Lambda^+_c\rightarrow nK^0_S\pi^+$ are
selected from the remaining tracks recoiling against the ST
$\bar{\Lambda}^-_c$ candidates.
A pion with charge opposite to the ST
$\bar{\Lambda}^-_c$ is selected, and a $K^0_S$ candidate is selected
with the same selection criteria as described above but without the $M_{\pi^+\pi^-}$ mass requirement.
If more than one $K^0_S$ candidate is formed, the one with the largest decay length significance $L/\sigma_L$ is retained,
where $\sigma_L$ is the vertex resolution of $L$.

Since the neutron is not detected, we use a kinematic variable
$$M^2_{\rm miss} \equiv E^2_{\rm miss}/c^4-|\overrightarrow{p}_{\rm miss}|^2/c^2$$
to obtain information on the missing neutron, where $E_{\rm
miss}$ and $\vec{p}_{\rm miss}$ are the missing energy and momentum
carried by the neutron, respectively, which are calculated by $E_{\rm
miss} \equiv E_{\rm beam}-E_{K^0_S}-E_{\pi^+}$ and $\vec{p}_{\rm
miss} \equiv \vec{p}_{\Lambda_c^+}-\vec{p}_{K^0_S}-\vec{p}_{\pi^+}$, where
$\vec{p}_{\Lambda_c^+}$ is the momentum of the $\Lambda_c^+$ baryon,
$E_{K^0_S}$ ($\vec{p}_{K^0_S}$) and $E_{\pi^+}$ ($\vec{p}_{\pi^+}$)
are the energies (momenta) of the $K^0_S$ and $\pi^+$, respectively.
Here, the momentum $\vec{p}_{\Lambda_c^+}$ is given by
$\vec{p}_{\Lambda_c^+}=-\hat{p}_{\rm tag}\sqrt{E_{\rm
beam}^2/c^2-m^2_{\bar{\Lambda}^-_c}c^2}$, where $\hat{p}_{\rm tag}$ is the
direction of the momentum of the ST $\bar{\Lambda}^-_c$ and
$m_{\bar{\Lambda}^-_c}$ is the nominal $\bar{\Lambda}^-_c$
mass~\cite{pdg2014}. If the $K^0_S$ and $\pi^+$ from the decay
$\Lambda^+_c\rightarrow n K^0_S\pi^+$ are correctly identified, the
$M^2_{\rm miss}$ is expected to peak around the nominal neutron mass
squared.

\begin{figure}[tp!]
\begin{center}
   \includegraphics[width=0.8\linewidth]{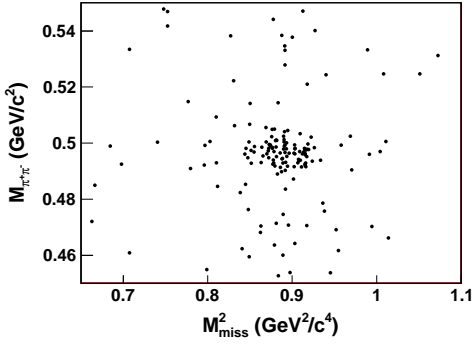}
   \caption{ Scatter plot of $M_{\pi^+\pi^-}$ versus $M^2_{\rm miss}$ for $\Lambda_c^+\rightarrow nK^0_S\pi^+$
   observed from data.}
\label{fig:scat_nkspi}
\end{center}
\end{figure}

The scatter plot of $M_{\pi^+\pi^-}$ versus $M^2_{\rm miss}$ for the $\Lambda^+_c\to
nK^0_S\pi^+$ candidates in data is shown in Fig.~\ref{fig:scat_nkspi}, where a cluster of events in the signal region is clearly visible.
According to MC simulations, the
dominant backgrounds are from the decays $\Lambda^+_c
\to \Sigma^-\pi^+\pi^+$ and $\Lambda^+_c \to \Sigma^+\pi^+\pi^-$ with $\Sigma^\pm\to n \pi^\pm$, which have
the same final state as signal.
These background events form a peaking background in $M^2_{\rm miss}$, but are distributed flat
in $M_{\pi^+\pi^-}$.
Backgrounds from non-$\Lambda_c^+$ decays are estimated by examining
the ST candidates in the $M_{\rm BC}$ sideband $(2.252, 2.272)$~GeV/$c^2$ in data,
whose area is 1.6 times larger than the background area in the signal region.

\begin{figure}[tp!]
\begin{center}
   \includegraphics[height=6cm,width=8.5cm]{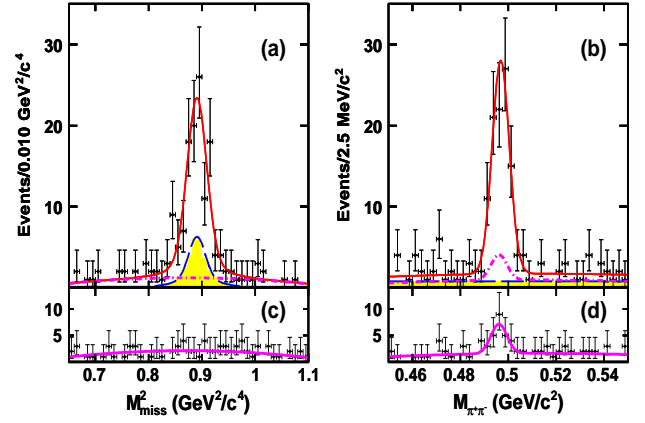}
   \caption{Simultaneous fit to $M^2_{\rm miss}$ and $M_{\pi^+\pi^-}$ of
   events in (a, b) the $\bar{\Lambda}_c^-$ signal region and (c, d) sideband
   regions. Data are shown as the dots with error bars. The
   long-dashed lines (blue) show the $\Lambda_c^+$ backgrounds while
   the dot-dashed curves (pink) show the non-$\Lambda_c^+$ backgrounds. The (red) solid curves show the total fit. The (yellow) shaded area show the MC simulated
   backgrounds from $\Lambda_c^+$ decay. } \label{fig:2Dfit_data}
\end{center}
\end{figure}

To obtain the yield of $\Lambda^+_c\to nK^0_S\pi^+$ events, we
perform a two-dimensional unbinned maximum likelihood fit to the $M^2_{\rm
miss}$ and $M_{\pi^+\pi^-}$ distributions in both $M_{\rm BC}$ signal and sideband regions simultaneously.
As verified with MC simulations, we model the $M_{\pi^+\pi^-}$ and
$M^2_{\rm miss}$ distributions with a product of two one-dimensional
probability density functions, one for each dimension.
The signal functions for $M^2_{\rm miss}$ and $M_{\pi^+\pi^-}$ are
both described by double Gaussian functions.
The peaking background in the $M^2_{\rm miss}$
distribution is described by a double Gaussian function with
parameters fixed according to MC simulations, and the flat distribution in the $M_{\pi^+\pi^-}$ spectrum is described by a constant function.
The non-$\Lambda_c^+$ decay background is modelled by a second-order
polynomial function in the $M^2_{\rm miss}$ distribution and a Gaussian
function plus a second-order polynomial function in the $M_{\pi^+\pi^-}$
distribution, in which the parameters and the normalized background yields are constrained by
the events in $M_{\rm BC}$ sideband in the simultaneous fit.
The fit procedure is validated by analyzing a large ensemble of MC-simulated samples,
in  which the pull distribution of the fitted yields is in good agreement with the normal distribution.
Projections of the final fit to data are shown in Fig.~\ref{fig:2Dfit_data}.
From the fit, we obtain $N^{\rm obs}_{nK^0_S\pi^+}=83.2\pm10.6$,
where the error is  statistical only.

The absolute branching fraction for $\Lambda_c^+\rightarrow
nK^0_S\pi^+$ is determined by
\begin{equation}
\mathcal{B}(\Lambda_c^+\rightarrow nK^0_S\pi^+)=\frac{N^{\rm
obs}_{nK^0_S\pi^+}}{N^{\rm
tot}_{\bar{\Lambda}_c^-}\times\varepsilon_{nK^0_S\pi^+}\times\mathcal{B}(K^0_S\rightarrow
\pi^+\pi^-)}, \label{eq:branch}
\end{equation}
where $\varepsilon_{nK^0_S\pi^+}$ is the detection efficiency for
the $\Lambda_c^+\rightarrow nK^0_S\pi^+$ decay, which does not
include the branching fraction for $K^0_S\rightarrow\pi^+\pi^-$.
For each ST mode $i$, the efficiency $\epsilon^i_{nK^0_S\pi^+}$ is
obtained by dividing the DT efficiency $\epsilon^i _{{\rm tag},nK^0_S\pi^+}$ by
the ST efficiency $\epsilon^i _{\rm tag}$.
Weighting $\epsilon^i_{nK^0_S\pi^+}$ by the ST yields in data for each tag mode,
we obtain $\varepsilon_{nK^0_S\pi^+}=(45.9\pm0.3)\%$.
Inserting the values of $N^{\rm obs}_{nK^0_S\pi^+}$, $N^{\rm tot}_{\bar{\Lambda}^-_c}$, $\varepsilon_{nK^0_S\pi^+}$ and
$\mathcal{B}(K^0_S\rightarrow\pi^+\pi^-)$~\cite{pdg2014} in
Eq.~(\ref{eq:branch}), we obtain $\mathcal{B}({\Lambda^+_{c}\rightarrow n K^0_S\pi^+})=(1.82\pm0.23)\%$,
where the statistical error, including those from $N^{\rm obs}_{nK^0_S\pi^+}$ and $N^{\rm tot}_{\bar{\Lambda}^-_c}$ is presented.

With the DT technique, the systematic uncertainties from the ST side cancel in the branching fraction measurement.
The systematic uncertainties for measuring
$\mathcal{B}(\Lambda_c^+\rightarrow nK^0_S\pi^+)$ mainly arise from
the uncertainties of PID, tracking, $K^0_S$ reconstruction and
the fit procedure. Throughout this paragraph, all quoted systematic
uncertainties are relative uncertainties.
The uncertainties in the $\pi$ PID and tracking are both
determined to be 1.0\% by studying a set of control samples of
$e^+e^-\rightarrow \pi^+\pi^-\pi^+\pi^-$, $e^+e^-\rightarrow
K^+K^-\pi^+\pi^-$ and $e^+e^-\rightarrow p\bar{p}\pi^+\pi^-$ based
on data taken at energies above 4.0 GeV. The uncertainty in the
efficiency of $K^0_S$ reconstruction is determined to be 1.5\% by
studying the control samples of $J/\psi\rightarrow K^{*\mp}K^{\pm}$
and $J/\psi\rightarrow \phi K^0_SK^{\pm}\pi^{\mp}$. The uncertainty
due to the fit procedure is estimated to be 5.2\% by varying the fit range,
the shapes of background and signal components, and the choice of
sideband regions. Besides these uncertainties mentioned above, there
are systematic uncertainties from the quoted branching fraction for
$K^0_S\rightarrow \pi^+\pi^-$ (0.1\%), the $N^{\rm tot}_{\bar{\Lambda}_c^-}$
(1.0\%) evaluated by using alternative signal shapes in fits to the
$M_{\rm BC}$ spectra, the MC statistics (0.6\%), the signal MC model
(1.3\%) estimated by taking into account the statistical variations
in the $M_{n\pi^+}$, $M_{nK^0_S}$ and $M_{K^0_S\pi^+}$ spectra
observed in data. These systematic uncertainties
are summarized in Table~\ref{tab:syst}, and the total systematic
error is estimated to be 5.9\% by adding up all the sources in
quadrature.

\begin{table}[tp!]
\caption{ Summary of the relative systematic uncertainties for $\mathcal{B}(\Lambda_c^+\rightarrow nK^0_S\pi^+)$.}
\begin{center}
\begin{tabular}
{lc} \hline \hline Source & Uncertainty \\ \hline
$\pi^{\pm}$ PID & 1.0\%  \\
$\pi^{\pm}$ tracking & 1.0\% \\
$K^0_S$ reconstruction & 1.5\% \\
Fit  & 5.2\% \\
$\mathcal{B}(K^0_S\rightarrow \pi^+\pi^-)$ & 0.1\% \\
$N^{\rm tot}_{\Bar{\Lambda}_c^-}$ & 1.0\% \\
MC statistics  & 0.6\% \\
MC Model   & 1.3\% \\
\hline Total  & 5.9\% \\
\hline \hline
\end{tabular}
\label{tab:syst}
\end{center}
\end{table}

In summary, using 567~pb$^{-1}$ of $e^+e^-$ collision data taken at
$\sqrt{s}=4.599$~GeV with the BESIII detector, we report the
observation of the decay $\Lambda^+_c\rightarrow nK^0_S\pi^+$. We measure the
absolute branching fraction for $\Lambda^+_c\rightarrow nK^0_S\pi^+$,
$\mathcal B({\Lambda^+_{c}\rightarrow n
K^0_S\pi^+})=(1.82\pm0.23\pm0.11)\%$, where the first uncertainty is
statistical and the second is systematic.
This is the first direct measurement
of a $\Lambda_c^+$ decay involving the
neutron in the final state since the discovery of the $\Lambda_c^+$ more than 30 years ago.
Quoting $\mathcal{B}(\Lambda_c^+\rightarrow pK^-\pi^+)$ and $\mathcal{B}(\Lambda_c^+\rightarrow pK_S^0\pi^0)$ measured by BESIII~\cite{1511.08380}, it can be found that the amplitudes of the above three decay processes satisfy the triangle relation and validate the isospin symmetry~\cite{1601.04241}.
Besides, we obtain $\mathcal{B}(\Lambda_c^+\rightarrow n\bar{K}^0\pi^+)/\mathcal{B}(\Lambda_c^+\rightarrow pK^-\pi^+)=0.62\pm0.09$ and $\mathcal{B}(\Lambda_c^+\rightarrow n\bar{K}^0\pi^+)/\mathcal{B}(\Lambda_c^+\to p\bar{K}^0\pi^0))=0.97\pm0.16$~\cite{brnk0bpi}, in which the common uncertainties have been cancelled in the calculation.
According to Ref.~\cite{1601.04241}, based on these ratios, the strong phase difference of $I^{(0)}$ and $I^{(1)}$  is calculated to be $\cos\delta=-0.24\pm0.08$, which is useful to understand the final state interactions in $\Lambda_c^+$ decays. Furthermore, the relative size of the two amplitudes $|I^{(1)}|/|I^{(0)}|$ is evaluated to be $1.14\pm0.11$, which indicates that the amplitude $I^{(1)}$ is not small as expected in the factorization scheme. This is consistent with the behaviors in the charmed meson decays~\cite{prd81_074021}. These results will be essential inputs for the study of other $\Lambda_c$ decays in theory.
Hence, the measurement of the neutron mode in this work provides the first complementary data
to the previously measured decays involving a proton,
which represents significant progress in studying the $\Lambda^+_c$.
The analysis method used in this work can also be extended to study more decay modes involving a neutron.

Lei Li, X.-R. Lyu and H.-L. Ma thank Wei Wang and Fu-Sheng Yu for useful discussions.
The BESIII collaboration thanks the staff of BEPCII and the IHEP
computing center for their strong support. This work is supported in
part by National Key Basic Research Program of China under Contract
No. 2015CB856700; National Natural Science Foundation of China (NSFC)
under Contracts Nos.\ 11235005, 11235011, 11275266, 11305090, 11305180, 11322544, 11335008, 11425524, 11505010; the
Chinese Academy of Sciences (CAS) Large-Scale Scientific Facility
Program; the CAS Center for Excellence in Particle Physics (CCEPP);
the Collaborative Innovation Center for Particles and Interactions
(CICPI); Joint Large-Scale Scientific Facility Funds of the NSFC and
CAS under Contracts Nos. U1232201, U1332201; CAS under Contracts
Nos. KJCX2-YW-N29, KJCX2-YW-N45; 100 Talents Program of CAS; National
1000 Talents Program of China; INPAC and Shanghai Key Laboratory for
Particle Physics and Cosmology; German Research Foundation DFG under
Contracts Nos. Collaborative Research Center CRC 1044, FOR 2359;
Istituto Nazionale di Fisica Nucleare, Italy; Joint Large-Scale
Scientific Facility Funds of the NSFC and CAS	under Contract
No. U1532257; Joint Large-Scale Scientific Facility Funds of the NSFC
and CAS under Contract No. U1532258; Koninklijke Nederlandse Akademie
van Wetenschappen (KNAW) under Contract No. 530-4CDP03; Ministry of
Development of Turkey under Contract No. DPT2006K-120470; NSFC under
Contract No. 11275266; The Swedish Resarch Council; U. S. Department
of Energy under Contracts Nos. DE-FG02-05ER41374, DE-SC-0010504,
DE-SC0012069, DESC0010118; U.S. National Science Foundation;
University of Groningen (RuG) and the Helmholtzzentrum fuer
Schwerionenforschung GmbH (GSI), Darmstadt; WCU Program of National
Research Foundation of Korea under Contract No. R32-2008-000-10155-0.
This paper is also supported by the Beijing municipal government under
Contract Nos.\ KM201610017009, 2015000020124G064.



\begin{thebibliography}{99}
\bibitem{brnk0bpi} In the calculation, we assume that the processes with $K^0_L$ and $K_S^0$ included have the same branching fractions.
\end{thebibliography}
\end{document}